\documentclass[12pt]{iopart}

 \usepackage{graphicx}

\begin{document}

\title[Broken $\mathcal{PT}$ symmetry]{An algebraically solvable $\mathcal{PT}$-symmetric potential with broken symmetry}
\author{E M Ferreira$^1$ and J Sesma$^2$}
\address{$^1$ Instituto de F\'{\i}sica, Universidade Federal do Rio
de Janeiro, 21941-972, Rio de Janeiro, Brasil}
\address{$^2$ Departamento de F\'{\i}sica Te\'orica, Facultad de
Ciencias, 50009, Zaragoza, Spain}

\eads{\mailto{erasmo@if.ufrj.br}, \mailto{javier@unizar.es}}

\begin{abstract}
The spectrum of a one-dimensional Hamiltonian with potential $V(x)=ix^2$ for negative real  $x$ and $V(x)=-ix^2$ for positive real $x$ is analyzed. The Schr\"odinger equation is algebraically solvable and the eigenvalues are obtained as the zeros of an expression explicitly given in terms of Gamma functions. The spectrum consists of one real eigenvalue and an infinite set of pairs of complex conjugate eigenvalues.
\end{abstract}

\pacno{03.65.Ge}



\nosections


The pioneering work of Bender and Boetcher \cite{ben1} marks the arising of $\mathcal{PT}$ symmetric quantum theory, a very active field of research which, besides having provided with amazing results in quantum mechanics, presents common aspects with other theories like integrable models or quantum fields. A tutorial introduction to the topic and a later complete update can be found in two papers by Bender \cite{ben2,ben3}, where numerous references to earlier developments are given. In recent years, three special issues of Journal of Physics A \cite{geye} have gathered articles dealing with different features of those non-Hermitian theories.

As one can learn from Ref. \cite{ben2}, a $\mathcal{PT}$-symmetric Hamiltonian $H$ may present broken or unbroken
symmetry. In the first case, complex eigenvalues appear besides of possible real ones. In the second case, the spectrum is purely real.
The necessary and sufficient conditions for the reality of energy eigenvalues have been set in the case of finite dimensional Hamiltonians \cite{ben4}. There have been also studies of that issue in the case of polynomial potentials depending on one or several parameters \cite{dela,shi1,shi2}. The reason why the symmetry becomes broken for certain sets of those parameters is understood, but a general explanation of the symmetry breaking mechanism is lacking. We believe that the study of different particular cases, as in Ref. \cite{leva}, may help to find a satisfactory theory.

Here we consider a parameterless $\mathcal{PT}$-symmetric Hamiltonian
\begin{equation}
H=-\,\frac{d^2}{dx^2}+V(x)  \label{e1}
\end{equation}
with
\begin{equation}
V(x)=\left\{\begin{array}{lll}i\,x^2,&\qquad \mbox{for}&\quad x\leq 0, \\ -\,i\,x^2, &\qquad \mbox{for}&\quad x\geq 0,\end{array}\right. \label{e2}
\end{equation}
where $\mathcal{PT}$-symmetry is broken. In fact, as we are going to show, only one of its eigenvalues is real; the rest of them form complex conjugate pairs.

The Schr\"odinger equation for the Hamiltonian (\ref{e1}) with potential (\ref{e2})
is algebraically solvable. A complex rotation of the variable by an angle of $\pi/8$ allows one to relate the problem at hand with that of resonances in a parabolic odd potential, discussed in a recent paper \cite{ferr} to which we refer for details of the solution of the Schr\"odinger equation. Normalizable wave functions are obtained for values of the energy $E$ satisfying the equation
\begin{equation}
\frac{e^{-i\pi/8}}{\Gamma\left(\frac{1-e^{i\pi/4}E}{4}\right)\,\Gamma\left(\frac{3-e^{-i\pi/4}E}{4}\right)}
+ \frac{e^{i\pi/8}}{\Gamma\left(\frac{1-e^{-i\pi/4}E}{4}\right)\,\Gamma\left(\frac{3-e^{i\pi/4}E}{4}\right)}=0\,.  \label{e3}
\end{equation}
It is immediate to see that if $E$ is a solution of this equation, its complex conjugate, $E^*$, is also a solution. Therefore, the eigenvalues of the Hamiltonian either are real or form complex conjugate pairs.

A glance at Eq. (38) of Ref. \cite{ferr} allows one to check that the eigenvalues of $H$ resulting from Eq. (\ref{e3}) coincide with the energies of the Gamow states discussed before  \cite{ferr} multiplied by $e^{i\pi/4}$, as was to be expected. We report, in Table 1, the lowest modulus solutions of Eq. (\ref{e3}). Figure 1 shows their positions in the complex energy plane.
\begin{table}
\centering
\caption{Approximate eigenvalues of the $\mathcal{PT}$-symmetric Hamiltonian with broken symmetry given in Eqs. (\ref{e1}) and (\ref{e2}).}
\begin{tabular}{rcr}
     \hline
$1.258091$ & &  \\
$4.991315$ & $\pm$ & $0.780486\;i$ \\
$8.618144$ & $\pm$ & $3.363257\;i$ \\
$11.85539$ & $\pm$ & $5.952472\;i$ \\
$14.97138$ & $\pm$ & $8.594195\;i$ \\
$18.02114$ & $\pm$ & $11.26933\;i$ \\
$21.02922$ & $\pm$ & $13.96756\;i$ \\
$24.00868$ & $\pm$ & $16.68275\;i$ \\
$26.96728$ & $\pm$ & $19.41094\;i$ \\
$29.91003$ & $\pm$ & $22.14940\;i$ \\
\hline
\end{tabular}
\end{table}
\begin{figure}
\begin{center}
\vspace{0.5cm}\includegraphics{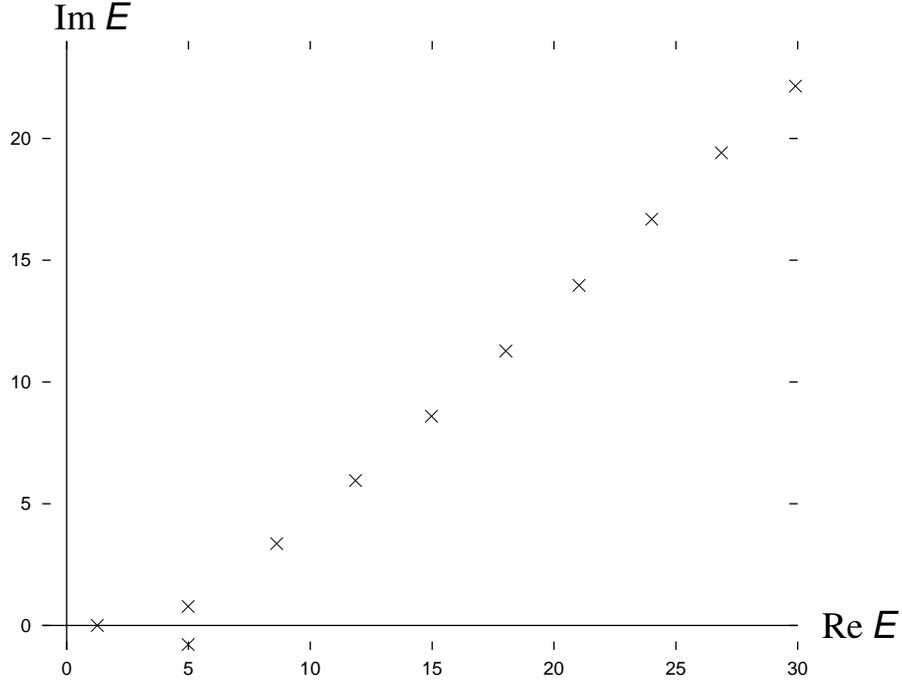}
\end{center}
\caption{Eigenstates of the Hamiltonian given in Eqs. (\ref{e1}) and (\ref{e2}). The graphic does not show the eigenstates on the lower half plane, located symmetrically to those on the upper half plane }
\label{fig:1}
\end{figure}
The corresponding eigenfunctions are
\begin{eqnarray}
\psi(x)=\mathcal{N}(E)\exp(-\alpha\,x^2/2)& \left[ b_1\ _1\!F_1\left(\frac{1-\alpha^{-1}E}{4};\frac{1}{2};\alpha\,x^2\right)\right. \nonumber\\
& \left.+\,b_2\,x\ _1\!F_1\left(\frac{3-\alpha^{-1}E}{4}; \frac{3}{2};\alpha\,x^2\right)\right],  \label{e4}
\end{eqnarray}
where $\mathcal{N}(E)$ represents a positive normalization constant (depending on the eigenvalue $E$) and we have used the abbreviation
\begin{equation}
\alpha = \left\{ \begin{array}{l} e^{i\pi/4} \quad \mbox{for}\quad x<0, \\ e^{-i\pi/4} \quad \mbox{for}\quad x>0. \end{array}\right.  \label{e5}
\end{equation}
Taking for the coefficients $b_1$ and $b_2$ the values
\begin{equation}
b_1=\frac{1}{\Gamma\left(\frac{1}{2}\right)}, \qquad b_2=\frac{-x}{|x|}\,\frac{\alpha^{1/2}\,\Gamma\left(\frac{3-\alpha^{-1}E}{4}\right)} {\Gamma\left(\frac{3}{2}\right)\,\Gamma\left(\frac{1-\alpha^{-1}E}{4}\right)}   \label{e6}
\end{equation}
guarantees the vanishing of $\psi(x)$ for $x\to\pm\infty$. (Notice that, in virtue of Eq. (\ref{e3}), $b_2$ takes the same value for $x<0$ and $x>0$.) In fact, for large (positive and negative) values of $x$,
\begin{equation}
\fl \psi(x)\sim\mathcal{N}(E)\,C\,\exp(-\alpha\,x^2/2)\left(\alpha\,x^2\right)^{-(1-\alpha^{-1}E)/4} \!\! \ _2\!F_0\left(\frac{1\! -\! \alpha^{-1}E}{4},\frac{3\! -\! \alpha^{-1}E}{4};;-\frac{\alpha^{-1}}{x^2}\right) ,  \label{e7}
\end{equation}
with the abbreviation
\begin{equation}
C=\alpha^{1-\alpha^{-1}E}\left[\frac{1}{\Gamma\left(\frac{1+\alpha^{-1}E}{4}\right)}-
\frac{\alpha^2\,\Gamma\left(\frac{3-\alpha^{-1}E}{4}\right)} {\Gamma\left(\frac{3+\alpha^{-1}E}{4}\right)\,\Gamma\left(\frac{1-\alpha^{-1}E}{4}\right)}\right],  \label{e8}
\end{equation}
and the values of $\alpha$ given in Eq. (\ref{e5}).

For illustration, we show the normalized wave functions of three eigenstates of the Hamiltonian,
corresponding to the eigenvalues 1.258091 (figure 2), 4.991315+0.780486$\,i$ (figure 3) and 8.618144+3.363257$\,i$ (figure 4). Our choice of the value of $b_1$ in Eq. (\ref{e6}) fixes the arbitrary phase of the wave function, making $\psi(x)$ to become real and positive at $x=0$. In the case of a real eigenvalue $E$,
$b_2$ turns out to be pure imaginary, and then
\begin{equation}
\psi (-x) = (\psi (x))^*,        \label{e9}
\end{equation}
a property that can be observed in figure 2 and checked in Eq (\ref{e4}). This equation proves also the relation existing between the wave functions $\psi_E(x)$ and $\psi_{E^*}(x)$ corresponding to complex conjugate eigenvalues, namely
\begin{equation}
\psi_E (-x) = (\psi_{E^*} (x))^*.        \label{e10}
\end{equation}
\begin{figure}
\begin{center}
\vspace{0.5cm}\includegraphics{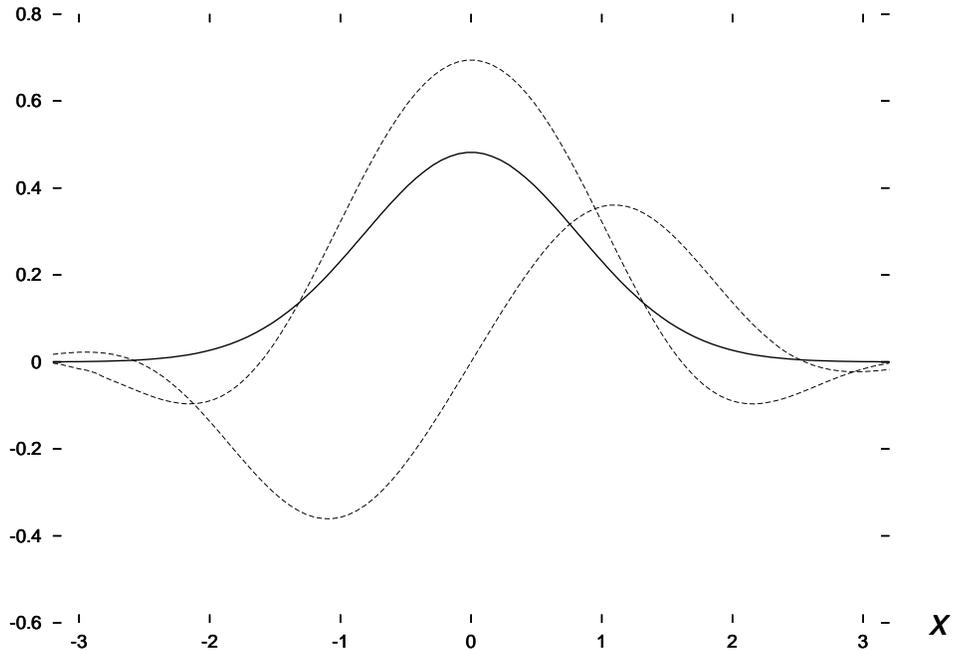}
\end{center}
\caption{Normalized wave function of the eigenstate of energy 1.258091. The solid line corresponds to the squared modulus, $|\psi(x)|^2$. The arbitrary phase of $\psi(x)$ has been chosen in such a way that $\psi(0)$ becomes real and positive. Then, the real and imaginary parts of $\psi(x)$ are represented by the dashed lines. }
\label{fig:2}
\end{figure}
\begin{figure}
\begin{center}
\vspace{0.5cm}\includegraphics{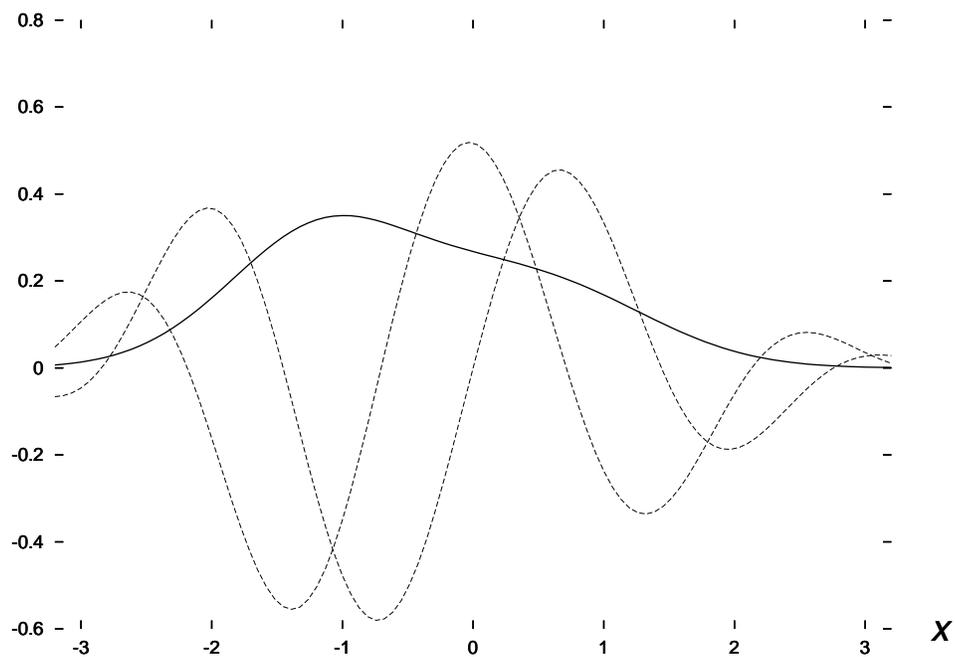}
\end{center}
\caption{Normalized wave function of the eigenstate of energy 4.991315+0.780486$\,i$. The comments in the caption of figure 2 concerning the solid and dashed lines are valid also here.}
\label{fig:3}
\end{figure}
\begin{figure}
\begin{center}
\vspace{0.5cm}\includegraphics{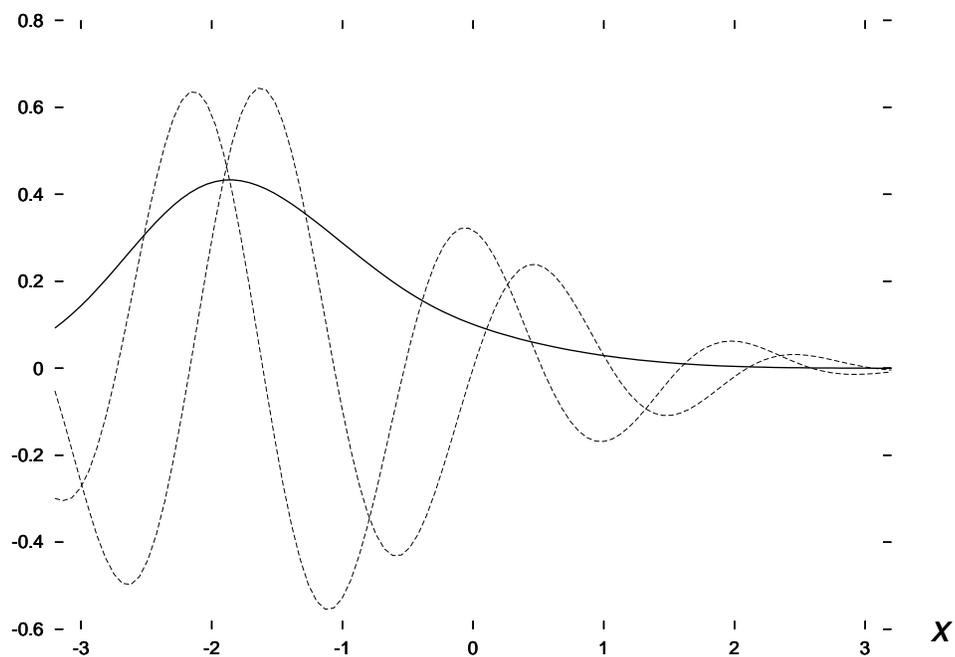}
\end{center}
\caption{Normalized wave function of the eigenstate of energy 8.618144+3.363257$\,i$. The conventions adopted in figures 2 and 3 are maintained here.}
\label{fig:4}
\end{figure}

Potentials of the form
\begin{equation}
V_a(x)=\left\{\begin{array}{lll}i\,(-x)^a,&\qquad \mbox{for}&\quad x\leq 0, \\ -\,i\,x^a, &\qquad \mbox{for}&\quad x\geq 0,\end{array}\right., \label{e11}
\end{equation}
with real $a$, are obviously $\mathcal{PT}$-symmetric. Their eigenvalues, with the boundary conditions on the real axis $\psi(\pm\infty+0i)=0$, are expected to be real. And this seems to be the case for $a\geq 3$, according to studies of Simon \cite{simo}, Caliceti {\em et al.} \cite{cali}  and Shin \cite{shi1}. Here we have considered the case $a=2$ and found only one real eigenvalue and a presumably infinite set of pairs of complex conjugate ones. Then, one may conjecture that, in the family of potentials (\ref{e11}), the $\mathcal{PT}$-symmetry is broken for $a<3$, similarly to what happens with the family of potentials \cite{ben2}
\begin{equation}
V_\beta(z)=\,-\,(iz)^\beta,   \qquad \beta\; {\rm real}\,,     \label{e15}
\end{equation}
with the boundary condition that
\[
\fl \psi(z)\to 0\; {\rm exponentially,\; as}\;z\to\infty\;{\rm along\; the\; two\; rays}\; \arg z = \left\{\begin{array}{ll} -\pi+\frac{\beta-2}{2\beta+4}\,\pi \\ -\frac{\beta-2}{2\beta+4}\,\pi\end{array}\right.\,,
\]
for $\beta<2$. Accordingly, pairs of real eigenvalues of the Hamiltonian with potential
(\ref{e11}) would turn progressively into complex conjugate pairs as $a$ decreases from 3 to 2.

\ack{
Financial support from Conselho Nacional de Desenvolvimento
Cient\'{\i}fico e Tecnol\'{o}gico  (CNPq, Brazil) and from Departamento de Ciencia, Tecnolog\'{\i}a y Universidad del Gobierno de Arag\'on (Project E24/1) and Ministerio de Ciencia e Innovaci\'on (Project MTM2009-11154) is gratefully acknowledged.}


\section*{References}

\begin{thebibliography}{99}

\bibitem{ben1} Bender C M and Boetcher S 1998 {\it Phys. Rev. Lett.} {\bf 80} 5243

\bibitem{ben2} Bender C M 2005 {\it Contemp. Phys.} {\bf 46} 277

\bibitem{ben3} Bender C M 2007 {\it Rep. Progr. Phys.} {\bf 70} 947

\bibitem{geye} Geyer H, Heiss D and Znojil M (ed) 2006 {\it J. Phys. A: Math. Gen.} {\bf 39} Number 32
    \nonum Fring A, Jones H F and Znojil M (ed) 2008 {\it J. Phys. A: Math. Theor.} {\bf 41} Number 24
    \nonum Bender C M, Fring A, G\"unther U and Jones H F 2012 {\it J. Phys. A: Math. Theor.} {\bf 45} Number 44

\bibitem{ben4} Bender C M and Mannheim P D 2010 {\it Phys. Lett. A} {\bf 374} 1616

\bibitem{dela} Delabaere E and Pham F 1998 {\it Phys. Lett A} {\bf 250} 29
    \nonum Delabaere E and Trinh D T 2000 {\it J. Phys. A: Math. Gen.} {\bf 33} 8771
    \nonum Bender C M, Berry M, Meisinger P N, Savage V M and Simsek M 2001 {\it J. Phys. A: Math. Gen.} {\bf 34} L31
    \nonum Handy C R 2001 {\it J. Phys. A: Math. Gen.} {\bf 34} 5065
    \nonum Handy C R, Khan D, Wang X Q and Tymczak C J 2001 {\it J. Phys. A: Math. Gen.} {\bf 34} 5593
    \nonum Dorey P, Dunning C and Tateo R 2001 {\it J. Phys. A: Math. Gen.} {\bf 34} 5679

\bibitem{shi1} Shin K C 2001 {\it J. Math. Phys.} {\bf 42} 2513
    \nonum Shin K C 2002 {\it Commun. Math. Phys.} {\bf 229} 543

\bibitem{shi2} Shin K C 2005 {\it J. Math. Phs.} {\bf 46} 082110
    \nonum Shin K C 2005 {\it J. Phys. A: Math. Gen.} {\bf 38} 6147

\bibitem{leva} L\'evai G 2012 {\it J. Phys. A: Math. Theor.} {\bf 45} 444020

\bibitem{ferr} Ferreira E M and Sesma J 2012 {\it J. Phys. A: Math. Theor.} {\bf 45} 415302

\bibitem{simo} Simon B 1970 {\it Ann. Phys. (N. Y.)} {\bf 58} 76

\bibitem{cali} Caliceti E, Graffi S and Maioli M 1980 {\it Commun. Math. Phys.} {\bf 75} 51

\end{thebibliography}
\end{document}